\documentclass{emulateapj}
\usepackage{color}
\usepackage{latexsym, graphicx, amssymb, longtable,epsf,ulem}
\def\g{\, {\rm g}}
\def\s{\, {\rm s}}

\newcommand{\x}{\sout}
\newcommand{\w}{ }
\newcommand{\rr}{\color{red}}
\newcommand{\z}{\color{blue}}

\newcommand{\abr}[1]{\textbf{\textit{AB: #1}}}

\newcommand{\icarus}{Icarus}

\long\def\symbolfootnote[#1]#2{\begingroup\def\thefootnote{\fnsymbol{footnote}}
\footnote[#1]{#2}\endgroup}

\shorttitle{Composition of the $\beta$\,Pic gas disk}
\shortauthors{Xie et al.}

\begin{document}

\title{On the {\w unusual gas} composition in the $\beta$ Pictoris debris
  disk}

\author{Ji-Wei Xie$^{1, 2}$, Alexis Brandeker$^3$, Yanqin Wu$^1$}
\affil{$^1$Department of Astronomy and Astrophysics, University of Toronto, Toronto, ON M5S 3H4, Canada; \\ jwxie@astro.utoronto.ca; wu@astro.utoronto.ca}
\affil{$^2$Department of Astronomy \& Key Laboratory of Modern Astronomy and Astrophysics in Ministry of Education, Nanjing University, 210093, China}
\affil{$^3$Department of Astronomy, Stockholm University, SE-106 91 Stockholm, Sweden; alexis@astro.su.se}



\begin{abstract}
  The metallic gas associated with the $\beta$\,Pic debris disk is
  {\rr not} believed {\rr to be} primordial, but arises {\rr from the}
  destruction of dust grains.  Recent observations have shown that
  carbon and oxygen in this gas are exceptionally overabundant
  compared to other elements, by some $400$ times. We study the origin
  of this enrichment under two opposing hypothesis, {\it preferential
    production}, where the gas is produced with the observed unusual
  abundance {\rr (as may happen if gas is produced by photo-desorption
    from C/O-rich icy grains)}, and \textit{preferential depletion},
  where the gas evolves to the observed state from an original solar
  abundance {\rr (if outgassing occurs under high-speed collisions)}
  under a number of dynamical processes.  We include in our study the
  following processes: radiative blow-out of metallic elements,
  dynamical coupling between different species, and viscous accretion
  onto the star.  We find that, if gas viscosity is sufficiently low
  (the conventional $\alpha$ parameter $\lesssim10^{-3}$),
  differential blow-out dominates. While gas accumulates gradually in
  the disks, metallic elements subject to strong radiation forces,
  such as Na and Fe, deplete more quickly than C and O, naturally
  leading to the observed overabundance of C and O.  On the other
  hand, if gas viscosity is high ($\alpha\gtrsim10^{-1}$, as expected
  for this largely ionized disk), gas is continuously produced and
  viscously accreted toward the star. This removal process does not
  discriminate between elements so the observed overabundance of C and
  O has to be explained by a {\it preferential production} that
  strongly favors C and O to other metallic elements. One such
  candidate is photo-desorption off the grains.  We compare our
  calculation against all observed elements ($\sim 10$) in the gas
  disk and find a mild preference for the second scenario, based on
  the abundance of Si alone. If true, $\beta$\,Pic should still be
  accreting at an observable rate, well after its primordial disk has
  disappeared.
\end{abstract}

\keywords{circumstellar matter Ñ planetary systems: formation Ñ
planetary systems: protoplanetary disks Ñ stars: individual ($\beta$ Pictoris)}


\section{Introduction}\label{SECN:INTRO}

The young \citep[$\sim$ 20\,Myr,][]{men08}, nearby
\citep[19\,pc,][]{van07} main-sequence (A5\,V) star $\beta$~Pictoris
has been known to harbour circumstellar (CS) gas ever since its
characterization as a ÔshellÕ star by \citet{sle75}. With the
discovery of CS dust by IRAS \citep{aum85} located in an edge-on disk
\citep{smi84}, the interest in the CS gas was renewed with more
detailed observations of the circumstellar absorption lines in both
the optical and the ultraviolet. 
The low estimated lifetime of CS dust
implies {\w that} it is probably replenished on short timescales
through destructive collisions, characteristic for a \textit{debris}
disk \citep{bac93}.

Early attempts to understand the properties and origin of the CS
\textit{gas} were hampered by the lack of constraints on what was
believed to be the dominant gas components: H, C, N, and
O. Observations of these elements are difficult because transition
lines are mostly in the far-UV, where the terrestrial atmosphere
blocks radiation and the stellar photospheric emission is very weak,
or in the infrared (IR) to far-IR for thermal emission from molecular
(H$_2$) or hyper-fine transitions (\ion{O}{1}, \ion{C}{2}), where
again spectroscopic observations from space are required. Another
problem was the unknown spatial distribution -- absorption
spectroscopy does not reveal \textit{where} along the line of sight
the gas is located, making solutions degenerate
\citep[e.g.][]{lag95}. In particular, the stellar radiation force on
e.g.\ \ion{Na}{1} and \ion{Ca}{2} is known to exceed gravity by more
than an order of magnitude, yet from the absorption line profile the
gas is observed to be at rest with respect to the star. The problem
was revisited when \citet{olo01} spatially resolved emission from
\ion{Na}{1} in orbit around the star, showing the gas to be widespread
in the disk. Something must prevent the gas from escaping on
free-ascend trajectories, but limits on the column density of H$_2$
\citep{lec01} together with limits on \ion{H}{1} \citep{fre95} shows
that H cannot be dense enough in the disk to significantly brake
{\w these metal species} \citep{bra04}. To solve this
enigma, a variety of braking scenarios were studied by \citet{fer06},
who found that an overabundance of C by a factor of at least 10 over
solar abundance would be
sufficient for charged ions in the disk to behave as a single fluid of
low effective radiation pressure {\w and
  remain} in orbit. C is an effective braking agent because of its
high abundance, low absorption coefficient in the optical (resulting
in a negligible radiation force), and an appreciable ionization
fraction. Since the observed elements experiencing a high radiation
force spend a tiny fraction of their time in {\w the} neutral
state, it is sufficient {\w that}
only ions are efficiently braked. Observations by the Far-UV Space
Explorer (FUSE) subsequently showed C to indeed be overabundant by
20\,$\times$ with respect to other species \citep{rob06}. Recently,
detailed absorption spectroscopy by HARPS at the ESO 3.6\,m telescope
\citep{bra11} and far-IR spectroscopy by Herschel/PACS of \ion{C}{2} 157.7\,$\mu$m
emission from $\beta$\,Pic \citep{bra12} have inferred an
even higher overabundance of both C and O, up to 400\,$\times$ the
solar ratio.

The large amounts of C in the CS gas explains why the metallic species
are not removed by the radiation pressure, but raises the new question
where this huge 400\,$\times$ overabundance comes from. \citet{fer06}
argued that the CS gas cannot be primordial, that is, a remnant from
the initial star-forming nebula. Instead, the gas is thought to be
secondary, {\w and is produced by the dust grains in the debris disk,
  either through} evaporation {\w during mutual collisions}
\citep[hereafter the ECG scenario,][]{cze07},{\w or through}
photo-desorption by UV radiation \citep[hereafter the PDG
scenario,][]{che07}.\footnote{The evaporation of comet-like
  bodies falling into the star \citep[e.g.][]{beu89} {\w has also been
    suggested to be responsible for producing gas very close to the
    star}.}  {\rr The two proposals produce gas with different
  chemical patterns: the former should lead to gas with
  solar-composition (hydrogen poor), the latter to C/O rich gas.  }

  If the observed abundances reflect those at production, these
  observations directly reveal the production mechanism, and more
  importantly, the chemical composition of the dust grains. The latter
  property is currently unconstrained: we do not yet know whether the
  grains are mostly icy or rocky, or if they are depleted in carbon
  like bodies in our Asteroid belt.  In the $\beta$\,Pic gas disk, not
  just C and O, but elements like Na, Mg, Al, Si, S, Ca, Cr, Mn, Fe,
  Ni are also observed. This affords us an unprecedented peek into the
  chemistry of planet formation in outer planetary systems.

  However, the current abundances of elements in the gas might not
  directly reflect those at production, as the abundances could have
  evolved in time. In particular, the various degrees of radiation
  force experienced by different species may result in some elements
  being removed at higher rates than others. Since the radiation force
  on both C and O is negligible, while it is significant on most other
  observed elements, this could potentially explain why C and O appear
  enriched with respect to other elements. In this paper, we {\w
    investigate the differential depletion of various elements in the
    $\beta$\,Pic disk, in an effort to explain the enrichment of C and
    O and to constrain the chemical composition of the dust grains. }
  In detail, our paper is organized as follows. {\w We describe our
    model in \S \ref{sec:model} with some of the detailed derivations
    delegated to Appendices A and B. The results are presented in \S
    \ref{subsec:result}. We discuss in \S \ref{sec:discussions} the
    implications of these results in the broad framework of planet
    formation and summarize our work in \S \ref{sec:summary}.}

\section{Model}
\label{sec:model}

We {\w study the evolution of CS gas using a simple one-zone model.}
{\w We consider} elements from {\rr Li to Ni with the exceptions of
  noble gases}. {\rr This gas is most likely second-generation gas,
  reproduced from solid materials after the primordial disk has
  dispersed. As such, the element hydrogen should be very
  underabundant, together with helium and other noble gases that do
  not easily condense.}  We include only the neutral and first ionized
states of the elements. {\rr Higher ionization states are negligible
  in the $\beta$ Pic disk because ionizations are mostly caused by the stellar
  photospheric spectrum, which has few very hard
  photons \citep{fer06, zag10}.}  {\w The element C, being abundant
  and significantly ionized, and experiencing little radiation
  pressure, is both the main donor of free electrons and the main
  braking agent (\ion{C}{2} ions) for metal species that are
  radiatively accelerated. In contrast, the other abundant element, O,
  remains largely neutral and is considered inert.}
We consider four processes: 
1)\,\textit{Gas production}, which we assume to be continuous and due
to {\w an unspecified} mechanism.
2)\,\textit{Outward radiation drift}. {\w Different gas species}
experience the radiation force to different degrees, parametrized by
the radiation force coefficient $\beta$ that is the ratio between the
radiation and gravity force. For $\beta>1$, radiation dominates over
gravity. {\rr But because of the centrifugal force, $\beta > 0.5$ is
  sufficient for particles born on circular orbits to be unbound from
  the star. }
3)\,\textit{Field particle interaction}. {\w Particles} that
{\w are accelerated} outward will interact with the field
particles, \ion{C}{2}, through Coulomb scattering.
4)\,\textit{Viscous accretion}. {\w Since the gas disk is
  significantly ionized, it is reasonable to assume that it is subject
  to the magnetorotational instability\citep{bal91, haw95}
  which allows the gas to viscously accrete into the star.}
In the following, we will investigate how these four
processes compete with each other to {\w determine the evolution of
  elemental abundances.}

{\rr We ignore the following processes: scattering between ions and
  neutral gas, and between ions and charged dust grains. Due to
  Coulomb focussing, ions like \ion{C}{2} are several orders of magnitude
  more efficient as a braking agent, when compared to
  neutrals.\footnote{ {\rr This applies even to braking atoms, as
      polarizability makes the ion-neutral cross-section several times
      larger than that of neutral-neutral interaction \citep{fer06}.}}
  Dust, while considered by \citet{fer06} as an effective braking
  agent, is much less efficient than \ion{C}{2} when the number density of
  \ion{C}{2} is as high as observed. }

The mass of the star is
$M_{\star}\sim \rm 1.75\,M_{\rm \odot}$ and the age of the system
$t_{\rm age}\sim20$\,Myr. {\w The radiation spectrum of $\beta$\,Pic is
  adopted from \citet{hau99}, and the radiation pressure is calculated
  following the procedure of  \citet{fer06} and \citet{zag10}. All elements
  are assumed to be in photo-ionization/recombination equilibrium.}
According to \citet{bra04}, gas (at least the element sodium)
follows a {\w spatial} distribution similar to {\w that of} the dust,
{\rm with most gas mass located at} $r \sim 100$\,AU {\w and extending
  outward to hundreds of AUs}.  {\w We simplify this into a one-zone model with
  {\rr $r=100$\,AU, }assuming a disk} temperature 
of $T_{\rm disk}\sim50$\,K {\w \citep{zag10}} and  {\w a}
carbon density {\w of} {\rr $N_{\rm C}= 50$\,cm$^{-3}$}{\w, as is
  appropriate for the mid-plane at {\rr 100\,AU}} \citep{bra11}.  

\begin{figure*}
\begin{center}
  \includegraphics[width=1.0\textwidth]{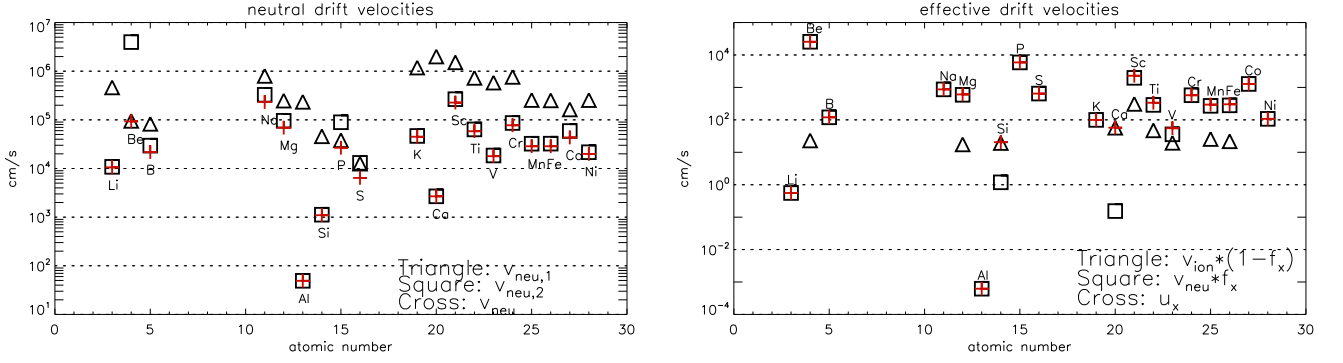}
  \caption{{\w Left} panel: {\w neutral drift velocities},
    $v_{\rm neu,{1}}$ (triangles), $v_{\rm neu,{2}}$ (squares) and
    $v_{\rm neu}$ (crosses), {\w plotted against atomic numbers for }
    elements with $\beta > 0.5$.  {\w Most neutral species satisfy $v_{\rm neu,{2}} < v_{\rm
        neu,{1}}$, or, drift velocities limited by the finite
      ionization time. } {\w Right} panel: {\w effective
      drift velocities $u_{\mathrm{x}}$ (red crosses), and respective
      contributions from the neutral state ($ f_{\mathrm{x}} v_{\rm
        neu}$) and from the first ionized state ($(1-f_{\mathrm{x}})
      v_{\rm ion}$), for elements that satisfy $\beta \geq 0.5$ in one
      or both of these states.  Drift in the neutral state dominate
      the effective drift, with the exception of elements Si and Ca.}
    Results {\w are based on}
    $N_{\mathrm{C}}={\rr 50}$\,cm$^{-3}$, $r={\rr 100}$\,AU, $T_{\rm
      disk}=50$\,K, $M_{\star}=1.75$\,M$_{\odot}$, and {\w the
      radiation spectrum of a $\beta$\,Pic-like star.}
  }
\label{v12i}
   \end{center}
\end{figure*}

\subsection{Radiative Acceleration and Collisional Braking }
A particle (ion or neutral atom) with $\beta>0.5$ is accelerated
outwards {\w by} the radiation force{\w, accompanied by random
  scattering via} collisions with field particles
\ion{C}{2}. Multiplying the radiative acceleration by the
mean-free-time after which the particle suffers a strong ($90\deg$
deflection) scattering, we obtain {\w its} equilibrium {\w outward}
drift velocity.  {\w Barring other processes, this velocity determines
  the gas removal rate from our one-zone model.}

{\w For an ion
  ``x'', this velocity is} (see Appendix B)
\begin{eqnarray}
v_{\rm ion} & \sim & 3.7 \beta \left[\frac{m_{\rm x}^{2}} {m_{\rm C}(m_{\rm C}+m_{\rm x})}\right]^{1/2} \left(\frac{N_{\rm C\,II}}{100\, \rm cm^{-3}}\right)^{-1/2}   \nonumber \\
& & \left(\frac{r}{100\, \rm AU}\right)^{-1} \left( \frac{M_{\star}}{\rm M_{\rm \odot}}\right)^{1/2} \left(\frac{T_{\rm disk}}{100 \, \rm K}\right) \,\, {\rm cm\,s^{-1}},
\label{vion}
\end{eqnarray} 
where $m_{\rm C}$ and $m_{\rm x}$ are the masses of a C atom and {\w the ion ``x''}, respectively, $r$ is the
radial distance to the central star of mass $M_{\star}$, $N_{\rm
  C\,II}$ is the number density of \ion{C}{2}, and $T_{\rm disk}$ is
the temperature of the gas disk.

For a neutral {\w atom} ``x'', the
equilibrium drift velocity is (see Appendix A)
\begin{eqnarray}
v_{\rm neu,{1}} & \sim & 80 \beta \left( \frac{m_{\rm x}}{m_{\rm C}} \right) \left(\frac{m_{\rm x}m_{\rm C}} {m_{\rm p}(m_{\rm x}+m_{\rm C})}\right)^{1/2}   \left( \frac{N_{\rm C\,II}}{100\, \rm cm^{-3}}\right)^{-1}  \nonumber \\
 & &  \left(\frac{r}{100\, \rm AU} \right)^{-2}  \left( \frac{M_{\star}}{\rm M_{\rm \odot}}\right) \left( \frac{P_{\mathrm{x}}}{10\, \rm \AA^{3}}\right)^{-1/2} \,  \rm cm\,s^{-1},
\label{vneu1}
\end{eqnarray}
where $m_{\rm p}$ is the mass of the proton, and $P_{\mathrm{x}}$
 the polarizability of neutral atom ``x''. {\w We adopt values
  from \citet{joh11} (http://cccbdb.nist.gov) and where unavailable we
  simply take it to be $P_{\mathrm{x}} =10^{-29} {\mathrm{m}}^{-3}$.}

 {\w The typical ionization time is short
  for neutrals in the $\beta$\,Pic disk. This warrants considering the
  limiting velocity an atom can be accelerated to before it becomes
  ionized:}
\begin{eqnarray}
v_{\rm neu,{2}}  \sim  \beta\frac{GM_{\star}}{r^{2}}\frac{1}{\Gamma}
 \sim  1000\beta\left(\frac{\Gamma_{\rm x,AU}}{10^{-7}\,\rm s^{-1}}\right)\, \rm cm\,s^{-1},
\label{vneu2}
\end{eqnarray}
where $G$ is the gravity constant, $\Gamma_{\mathrm{x,AU}}$ is the
ionization rate of the neutral atom ``x'' at $r=1$\,AU.  Following
\citet{bra11}\footnote{Note that we use a different naming convention
  compared to \citet{bra11}. The two neutral limiting velocities
  $v_{\rm neu,1}$ and $v_{\rm neu,2}$ correspond to $v_{\rm drift}$
  and $v_{\rm ion}$ in \citet{bra11}, while $v_{\rm ion}$ in this
  paper denotes the limiting velocity of the ionized tracer
  particles.},
{\w we adopt the following to be} the {\w actual drift}
velocity of the {\w atom,}
\begin{equation}
v_{\rm neu} = \frac{\gamma \, v_{\rm neu,{2}} }  {\gamma+1}, 
\label{vneu}
\end{equation}
where $\gamma=v_{\rm neu,{1}}/v_{\rm neu,{2}}$.

Since any given element switches between  {\w the} neutral and
{\w the} ionized state{\w s,} we combine $v_{\mathrm{neu}}$ and
$v_{\mathrm{ion}}$ to {\w obtain} the effective drift velocity
{\w for} element ``x'' as
\begin{eqnarray}
u_{\rm x}=f_{\rm x}{v_{\rm neu}}+(1-f_{\rm x})v_{\rm ion},
\label{ux}
\end{eqnarray}
where $f_{\rm x}$ is the neutral fraction of element ``x'' that quantifies the fraction of time that
is spent in the neutral state.

Fig.~\ref{v12i} shows the calculated $v_{\rm neu,{1}}, v_{\rm
  neu,{2}}, v_{\rm neu}$ and $v_{\rm ion}$ for all elements with
$\beta > 0.5$. We observe that:
\begin{enumerate}
\item most
  elements' neutral drift velocities $v_{\rm neu}$ are close to
  $v_{\rm neu,2}$ (the ionization limit), except for Be and P{\rr ,} {\w which
are less rapidly ionized.}
\item neutral
  drift ($v_{\rm neu}f_{\rm x}$) dominates the final effective outward
  drift for most elements, except for Si and Ca. {\w The ion drift
is almost completely suppressed by the strong coupling with \ion{C}{2}.}
\end{enumerate}

\subsection{Radiative Acceleration versus Viscous Accretion} 
{\w The outward drift of
  radiatively accelerated metals is moderated by the inward diffusion
  in the viscous accretion disk. The latter arises in our partially
  ionized disk since the strong coupling between the magnetic field
  and the gas allows the onset of magnetorotational instability
  \citep{bal91}. This instability leads to angular momentum
  transport to which one might ascribe a $\alpha$-like turbulent
  diffusivity \citep{sha73} }
\begin{equation}
\nu=\alpha v_{\rm s}h \sim \alpha \Omega_{\rm k} h^{2} 
\label{alpha}
\end{equation}
where $v_{\rm s}$ is the local sound speed, $h$ the local scale height
and $\Omega_{\rm K}$ the Keplerian angular frequency. {\w We set} $h
=0.033 \rm AU (r/ \rm AU)^{5/4}$ {\w and estimate} the typical
timescale{\w s} for viscous accretion  and 
radiation driven drift {\w to be}
\begin{eqnarray}
\label{tv}
t_{\rm vis} &  \sim &  \frac{r^{2}} {\nu} 
\sim   1.4\times10^{5} \,  \left(\frac{\alpha}{0.1}\right)^{-1}  \left(\frac{r} {\rm 100 \,AU}\right) \, \rm yr \\ 
\nonumber \\
t_{\rm drift}  & \sim  & \frac{r}{u_{\rm x}} 
\sim     4.7\times10^{5}\,  \left(\frac{u_{\rm x}}{100 \, \rm cm\,s^{-1}}\right)^{-1} \left(\frac{r}{\rm 100 \,AU}\right) \, \rm yr. 
\label{tv-tr}
\end{eqnarray}

{\w Note that $t_{\rm vis}$ is the same for all elements, while
  $t_{\rm drift}$ is specific to element x.}  Equating $t_{\rm vis}$
with $t_{\rm drift}$ we find that the critical $\alpha$, at which the
inward flow driven by viscous accretion is comparable to the outward
flow driven by the radiation force, is given by
 \begin{eqnarray}
 \alpha_{\rm cr}\sim 0.3\,\left(\frac{u_{\rm x}}{10^{3} \rm \,cm\,s^{-1}}\right).
 \end{eqnarray}
 As {\w is} shown in the {\w right} panel of
 Fig.~\ref{v12i}, many elements {\w (Na included)}
 have $u_{\rm x} {\w \sim 10^3} \rm \,cm\,s^{-1}$. {\w So unless $\alpha
 \ll 10^{-1}$, the inward viscous diffusion can compete against the
 outward radiative drift which differentially depletes element x.}

\subsection{{\rr Evolution of the Abundances}}

The key quantity out of this study is the abundance ratio.  We define
an abundance ratio between carbon and element x as one that is
normalized to the solar value \citep{and89},
$[\mathrm{C}/\mathrm{x}]={\rm log_{10}}(N_{\rm C}/N_{\rm x})-{\rm
  log_{10}}(N_{\rm C}/N_{\rm x})_{\rm Solar}$.  {\rr The observed
  value is $N_{\rm C}/N_{\rm Na}\approx 400(N_{\rm C}/N_{\rm Na})_{\rm
    Solar}$ or $[\rm {C}/{Na}] \approx 2.6$ . It is the goal of this
  paper to understand how this large factor comes about}. 

We study a model where {\rr the gas is} continuously produced at a
constant rate, $S_C$, $S_x$, and with a fixed abundance ratio
[C/x]$_{\rm src}$. The value of [C/x] in the CS gas evolve in time as
the element x is subject to differential removal by radiation pressure
while all elements viscously diffuse to the star at the same rate. The
evolution is described by the following equations
\begin{eqnarray}
{{dN_{\rm c}}\over{dt}} & = & S_c - {{N_c}\over{t_{\rm vis}}},\nonumber \\
{{dN_{\rm x}}\over{dt}} & = & S_x - {{N_x}\over{t_{\rm vis}}} - {{N_x}\over{t_{\rm drift}}},
\label{eq:evolution}
\end{eqnarray}
{\rr which have solutions
\begin{eqnarray}
N_{\rm c} & = & S_{\rm c} t_{\rm vis} (1-e^{-t/t_{\rm vis}}) ,\nonumber \\
N_{\rm x} & = & S_{\rm x} t_{\rm eff} (1-e^{-t/t_{\rm eff}}) ,
\label{eq:solve}
\end{eqnarray}
where $t_{\rm eff} = t_{\rm vis}t_{\rm drift} / (t_{\rm vis}+t_{\rm drift})$. }
{\rr Therefore, the evolution of the abundance ratio is
\begin{eqnarray}
\left[\mathrm{C}/\mathrm{x}\right] =   \left[\mathrm{C}/\mathrm{x}\right]_{\rm src} + {\rm log}_{\rm 10} \left[\left(\frac {t_{\rm vis} } { t_{\rm eff}}\right) \left(\frac{1-e^{-t/t_{\rm vis}}}{1-e^{-t/t_{\rm eff}}}\right) \right].
\label{ratio}
\end{eqnarray} }
Elements that do not experience significant
radiation pressure are easily incorporated into the above analysis
by setting $t_{\rm drift} \rightarrow \infty$.


\subsection{Results}
\label{subsec:result}

\begin{figure}[h]
\begin{center}
  \includegraphics[width=0.5\textwidth]{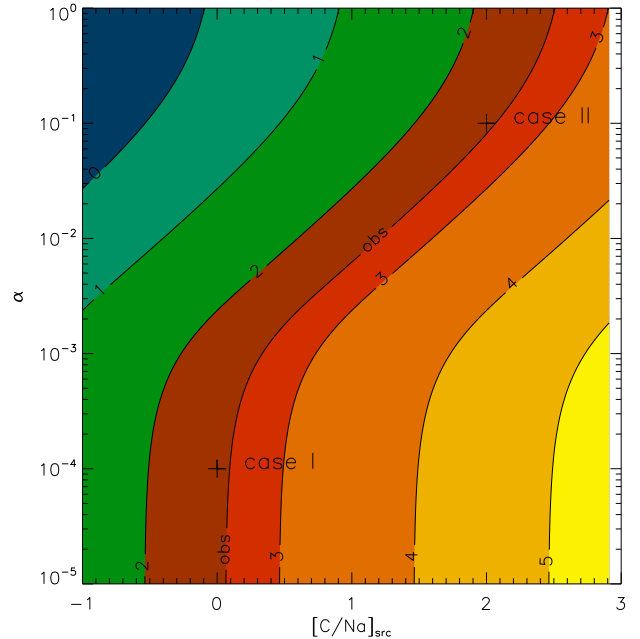}
  \caption{Mapping the{\w present} abundance ratio
    $[\mathrm{C}/\mathrm{Na}]$ (normalized to the solar abundance with
    logarithm scale) in the $\alpha-[\mathrm{C}/\mathrm{Na}]_{\rm
      src}$ plane. The contour line marked with ``obs'' denotes the
    abundance ratio of carbon and sodium
    ($[\mathrm{C}/\mathrm{Na}]=2.6$) that is about 400 times enhanced
    with respect to the solar abundance, as suggested by
    \citet{bra11}. {\w As expected, the present
      $[\mathrm{C}/\mathrm{Na}]$ value increases with
      $[\mathrm{C}/\mathrm{Na}]_{\rm src}$ and decreases for larger
      $\alpha$. At small viscosity, {\rr e.g., $\alpha < 10^{-4}$},
      the viscous timescale is longer than the current age and
      $[\mathrm{C}/\mathrm{Na}]$ becomes independent of $\alpha$. }
    {\w In} order to {\w re}produce the observed
    $[\mathrm{C}/\mathrm{Na}]$ {\w value}, the abundance of carbon in
    the gas source should be either near solar with low viscosity
    ($\alpha < 10^{-3}$) or super-solar with high viscosity ($\alpha >
    10^{-3}$); a sub-solar abundance case can be excluded. {\rr Two
      crosses mark a typical solar (case I) and a typical super-solar
      cases (case II) that both reproduce the observed carbon
      enhancement. In Fig.~4, we study these cases in more detail.}}
    \label{alpha-abund}
   \end{center}
\end{figure}

{\w We first focus on Na, the element with most detailed observations
  \citep{bra04, bra11, bra12}.\footnote{{\rr While Na in the $\beta$
      Pic disk mostly exists in ionized form that remains invisible to
      us, extensive modelling efforts \citep{fer06, zag10} have led to
      estimate of the electron density (and therefore \ion{Na}{2}/\ion{Na}{1} ratio)
      that consistently reproduce the observed ionization fractions in
      various elements. These are listed in Table
      \ref{tab:ionfrac}.}}}  Our calculations indicate that there are
two possibile scenario to reproduce the observed value of {\rr [C/Na]
  $\approx 2.6$}. One is that the metallic gas is produced at the
solar-composition, [C/Na]$_{\rm src} \sim 0$, but experiences a
negligible turbulent viscosity, $\alpha \leq 10^{-3}$. We call this
{\it `preferential depletion' }.  The other is that the gas is
produced with enriched C (and O) abundances, [C/Na]$_{\rm src} > 0$,
and it is advected into the star with larger $\alpha$. We call this
{\it `preferential production'}.  Our calculation places a constraint
on the relationship between $\alpha$ and [C/Na]$_{\rm src}$, as shown
in Fig.~\ref{alpha-abund}. But short of other theoretical or
observational evidences, both scenarios can explain the observed C/Na
overabundance.

\begin{figure*}
\begin{center}
\includegraphics[width=\textwidth]{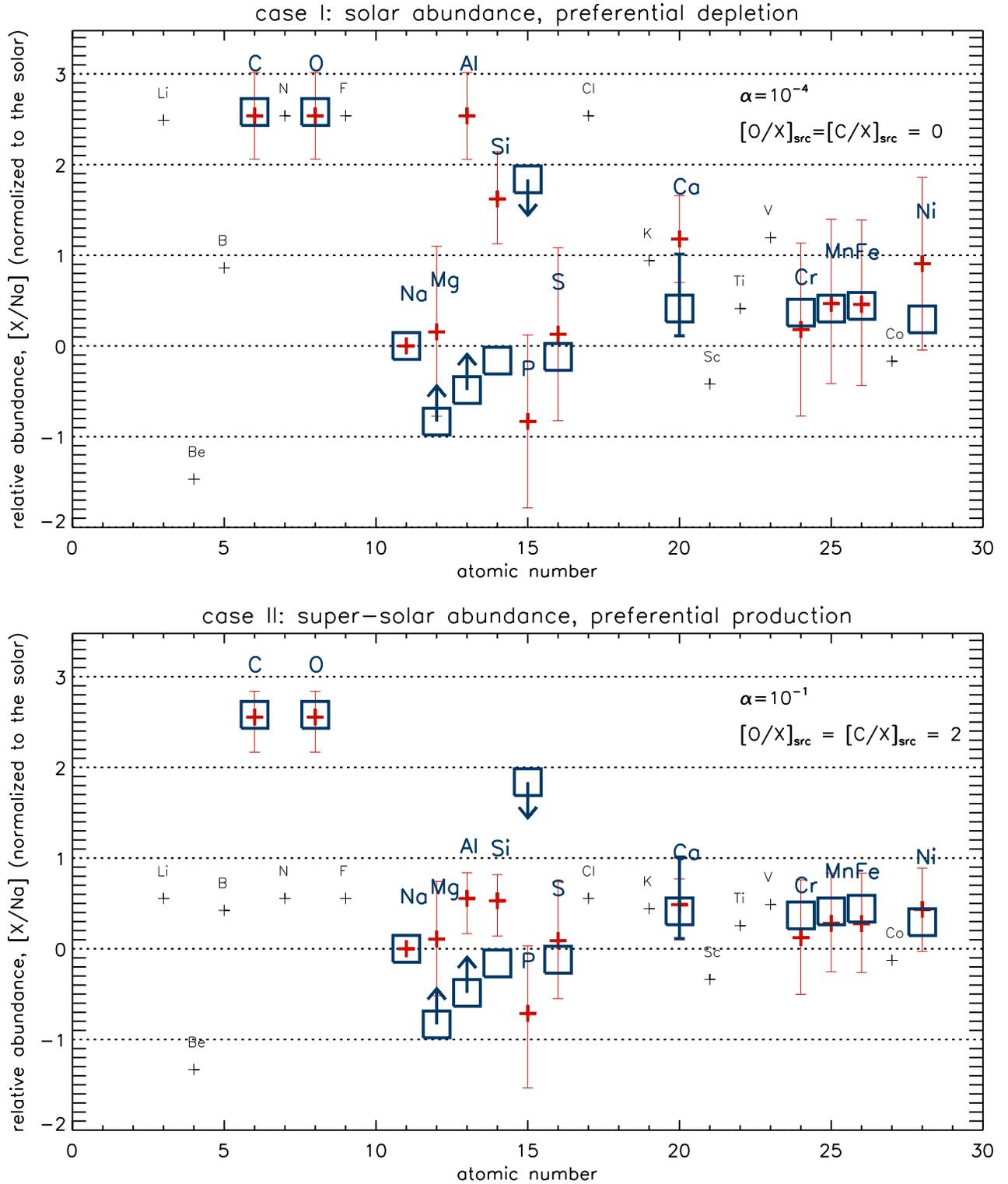}
\caption {{\w Abundance pattern in the $\beta$\,Pic disk, normalized to
    the solar value, and with all elements scaled by the element Na.
    The blue boxes, arrows, and errorbars represent the observed values and the red crosses
    the results of our calculations {\rr with the vertical red errorbars indicating effects of one source of theoretical uncertainties, when we vary the ionization fraction by a factor of 3 each way from those calculated in table 1.} The top panel shows the case when
    all elements are produced at the solar pattern, and the viscosity
    is low ($\alpha=10^{-4}$); the bottom panel shows the case when
    all elements are produced at the solar pattern except for C and O
    which are produced $100$ times more abundantly, and when $\alpha =
    0.1$. The size of the blue box reflects a typical uncertainty of
    $40\%$ \citep{lag98}, and the arrows indicate upper limits
    (for P) and lower limits (for Mg and Al which have saturated
    absorption lines), respectively.}  {\w Calculated
    values for elements that are as yet undetected are plotted as
    small black crosses. Also see Table \ref{tab:ionfrac} for tabulated
    values, {\rr as well as references for the observational ones}.} 
    }
\label{abund1}
   \end{center}
\end{figure*}

Now we turn to examine other metallic species in the $\beta$\,Pic
  disk, in the hope of extracting further constraints beyond what is
  achieved using Na alone. We study two specific sets of parameters,
  both reproducing {\rr [C/Na] $\approx2.6$ } at current time: solar
  abundance production and $\alpha=10^{-4}$ (preferential depletion);
  and super-solar production, [C/x]$_{\rm src} = 2$ and $\alpha=0.1$
  (preferential production). We assume for both cases that [C/O] = 0,
  or Oxygen is produced in tandem with carbon. This is based on the
  considerations that O is similar to C in dynamics (experiencing only
  viscous diffusion), and that the observed [C/O] $\approx 0$ {\rr
    \citep{bra12} }. The resultant [C/x] are plotted in
  Fig.\ref{abund1}, and summarized in Table \ref{tab:ionfrac}, {\rr
    for all elements between Li and Ni, with the exceptions of noble
    gases.}

{\w In the case of solar production (top panel of Fig.~\ref{abund1}),
  the present day elemental abundances are determined by their
  effective drift velocities (right panel of Fig.~\ref{v12i}), with
  elements that experience strong radiation drift to be severely
  depleted relative to C and O. This is the case for Na, Mg,P, S, Cr,
  Mn, Fe, Ni.\footnote{The CS gas of $\beta$\,Pic is likely of
    secondary origin, so a lower than solar abundance in H is not
    surprising.} But we expect elements that are not radiatively
  accelerated, like Al, and to some degree, Si, to be undepleted
  relative to C and O. This expectation contrasts strongly against the
  observations of Si, which is detected at a value $100$ times lower
  than our model prediction.\footnote{\rr{This conclusion is not significantly altered when varying $\alpha$.}}
  This difference is not explainable by hiding most Si in higher
  ionization states: in the $\beta$\,Pic disk, Si should be mostly in
  the first ionized state, which is easily observable. The observation
  of Al is more ambiguous. \citet{lag98} only reported a lower
  limit to Al since the \ion{Al}{2} line at 1670.79\,\AA\ is very
  saturated. Based on the observed line profile, a rough
  back-of-the-envelope estimate yields that Al is likely $\sim 10^2$
  more abundant than the lower limit. If true, this would place the
  observed Al abundance in between values expected for the two
  opposing scenario. The abundance of Al being crucial for deciphering
  the gas origin, a re-analysis of its line profile seems warranted.}

{\w The agreement between theory and observation is mildly improved in
  the case where the gas production is super-solar (in C and O), and
  where gas removal is dominated by turbulent diffusion (bottom panel
  of Fig.~\ref{abund1}).  The discrepancy, present still in Si, is a
  factor of $10$ or smaller. {\rr For instance, for a production abundance of [C/x] = 3, the predicted Si is $\sim5$ times higher than the observed value.}  In the following, we discuss the
  implications of these results.}

\begin{table*}
  \caption{{\w Column densities and Abundunces: observation vs.
      theory}$^{a}$
\label{tab:ionfrac}}
  \begin{tabular}{@{}ccccc|ccccc@{}}
  \hline
  \tableline \\
  atomic  & {\w element}& {\w neutral column}  
 &  {\w ion column}  
&{\w $f_x$} &  {\w $f_x$}& {\w [x/Na]} & {\w [x/Na]} &{\w References} \\ 
  number &&  {\w density}   
&  {\w density} & {\w (obs)$^{c}$}   & {\w (cal)$^{c}$}   & {\w (obs)} & {\w (cal)$^{d}$} & \\ 
  & &{\w [cm$^{-2}$]}$^{b}$ &{\w [cm$^{-2}$]$^{b}$} & & & & &
  \\  \tableline
%
6  & C&$1.4\times10^{17}$&$2.9\times10^{17}$&0.33 & 0.52   & 2.58                 & {\rr 2.54}; 2.55       &[1]  \\[2pt]
8  & O&$ 8.6 \times 10^{17}$   & -  & - & $\sim1$   & 2.58                    & {\rr 2.54}; 2.55  &[1]  \\[2pt]
11 & Na&$(3.4\pm0.4)\times 10^{10}$ &  $[9.2\times10^{12}]$ &-&$3.7\times10^{-3}$ & 0  & 0; 0   & [2]   \\[2pt]
12 & Mg &  $2.5\times10^{11}$ &$>2.4\times10^{13}$& $  <1\times10^{-2}$ & $8.6\times10^{-3}$   & $>-0.83$           & 0.15; 0.11        &[3], [4] \\[2pt]
13 & Al &  $<2.7\times10^{12}$&$>4.4\times10^{12}$& $<0.4$ & $1.3\times10^{-5}$ & $>-0.49$                 & {\rr 2.54}; 0.55        &[3]  \\[2pt]
14 & Si&$<1.1\times10^{13}$&$1.1\times10^{14}$ & $  <1\times10^{-1}$& $1.1\times10^{-3}$ &-0.16                 & 1.62; 0.53         &[3] \\[2pt]
15 & P & $<7\times10^{11}$ & $< 9.2\times10^{13}$  & - &0.22   &$<1.8$                   & -0.83; -0.71        &[5] \\[2pt]
16 & S& $5.4\times10^{12}$ & $[4.9\times10^{13}]$ & - &0.1   &-0.12                  & 0.13; 0.09         &[3] \\[2pt]
20 & Ca&  $<2\times10^{9}$&$2.6^{+7.6}_{-1.2}\times10^{13}$ &   $  <1.4\times10^{-4}$& $5.7\times10^{-5}$&$0.41^{+0.6}_{-0.3}$&1.18; 0.49      &[3],[6],[7] \\[2pt]
24 & Cr & $[3.5\times10^{10}]$ &   $4.8\times10^{12}$ & -& $7.4\times10^{-3}$&   0.37                    & 0.18; 0.12         &[3] \\[2pt]   
25 & Mn & $<2.4\times10^{10}$&$3.8\times10^{12}$& $  <6\times10^{-3}$& $9.4\times10^{-3}$&  0.41                    &  0.47; 0.28         & [3]  \\[2pt]
26 & Fe &  $4.9\times10^{11}$&${\rr 1.6}\times 10^{14}$& $  {\rr3.1}\times 10^{-3}$ &   $9.8\times10^{-3}$ & 0.44                  &  0.46; 0.28         &[3], [7]   \\[2pt]
28 & Ni &  $<7.6\times10^{10}$ &$1.5\times10^{13}$& $ <5\times10^{-3}$&$5.4\times 10^{-3}$ &  0.29                  &  0.91; 0.43           &[3], [7]   \\[2pt]
  \hline
\tableline 
\end{tabular}
\tablenotemark{References: [1] \citet{bra11} [2] \citet{bra04}, [3] \citet{lag98}, [4] \citet{vid94}}\\
\tablenotemark{\ \ \ \ \ \ \ \ \ \ \ \ \ \ \ \  [5]\citet{rob06}, [6] \citet{vid86}, [7] \citet{cra94}} \\
\tablecomments{$^{a}$ {\w Only elements that have reported column
    densities (for either neutral or first ionized state) are
    presented here.}}  \tablecomments{$^{b}$ {\w Observed values for
    column densities from absorption line studies (except C/O/Na where
    emission lines are detected).  Most error bars are of order $
    40\%$ of the measured values \citep{lag98}, unless explicitly
    listed.  Where multiple absorption features from the same lower
    state are present for the same species, we adopt the largest value
    of column density. If the lower states are different, we add up
    the column densities.  Values in square brackets are calculated
    assuming ionization equilibrium ($f_x$ cal).}}
\tablecomments{$^{c}$ {\w Neutral fraction}}
\tablecomments{$^{d}$ Abundances relative to solar, normalized by
  Na. Values in front of the semicolons are for the solar case, while
  those behind are for the super solar case (top and bottom panels of
  Fig.~\ref{abund1}, respectively).}
\end{table*}



\section{Discussion}
\label{sec:discussions}

{\w Here, we consider other astrophysical constraints on our scenario,
  and what future observations may bear on our current
  deliberation. The implications on the process of gas generation, as
  well as how our results are projected in the broad picture of planet
  formation, are also discussed here.}
   
\subsection{Viscosity in the gas disk}
\label{subsec:alpha}

Since the gas disk has a high ionization fraction ($\sim 20\%$), one expects it to be subject to the
magnetorotational instability (MRI), which has been shown numerically
to lead to turbulent viscosity of order $\alpha \sim1 0^{-2}$ or
higher \citep{haw95}.  {\w We consider briefly the complication due to
  the presence of charged dust grains in the debris disk. These grains
  couple strongly to the charged particles \citep{fer06}. The
  mean-free-time for electron-dust scattering is longer than the
  typical electron gyration time, so the electrons are expected to be
  tightly coupled to the magnetic field.  The grains, however, may
  substantially enhance the resistivity. The latter effect is known to
  reduce $\alpha$ \citep{bal09}, down to values of order $10^{-4}$ or
  lower \citep{fle95, bai11}. A more detailed study on MRI in the
  presence of charged dust is required to determine the realistic
  $\alpha$ value. }
 
\subsection{The rate of gas production and accretion}

{\w We obtain the current total mass of the gas disk by combining the
  observed column densities of C in Table \ref{tab:ionfrac} with an assumed scale height $h\sim0.1
  r$. Further assuming that C and O dominate the total gas mass
    and are produced at the solar ratio relative to each other, we
  estimate the required gas production rate to be}
\begin{eqnarray}
  {\w {{dM}\over{dt}}}
  \sim \left\{ \begin{array}{ll}
       1.2\times10^{13}\left(\frac{\alpha}{0.1}\right) \rm\,g\,s^{-1}  \  & \textrm{if $\alpha\gtrsim7\times 10^{-4}$}\\ \\
      8\times10^{10}\rm\,g\,s^{-1} \  & \textrm{otherwise.}\\ 
\end{array} \right.
\label{pgas}
\end{eqnarray}
We compare these values against various models for gas production.
{\w For the solar abundance case, \citet{cze07} estimated the gas
  production rate in their ECG model as} $5\times10^{11}$\,g\,s$^{-1}$
for silicate grains and $2\times10^{13}$\,g\,s$^{-1}$ for icy grains,
in the $\beta$\,Pic disk. {\w These are roughly consistent with our above
requirement.}

{\w Photo-desorption is a class of model where molecules and atoms are
  ejected from grains by UV photons.  \citet{che07} estimated the
  photo-desorption rate for Na, finding a Na production rate of $\sim
  1.33 \times 10^{33}$\,s$^{-1}$ for $\beta$\,Pic. This corresponds to
  a total gas production rate of $\sim 1.0 \times 10^{13} \g \s^{-1}$
  if photo-desorption efficiency is the same for C, N and O, as for
  Na.  \citet{obe07} performed experiments of photo-desorption of CO
  and CO$_2$ gas from icy grains. They found rates of order $10^{-3}$
  molecules per UV photon, or, $\sim 100$ times higher than, e.g.,
  those used by \citet{che07} for Na.  So if the dust grains have
  solar abundance and if all C and O are stored in the form of CO or
  CO$_2$ ice, photo-desorption would naturally yield a gas that is
  enriched in C and O by $\sim 100$ times relative to Na (and perhaps
  other elements).  Moreover, the gas production rate would be $\sim
  10^{15} \g \s^{-1}$.}
{\w So interestingly, both classes of models have no difficulty in
delivering the required amount of gas. We cannot exclude either model
using this discriminant.}

The dust disk is also losing grains through grinding down and
radiation pressure push-out. {\rr Taking a vertical optical depth of
  $10^{-3}$ for the grains at 5\,$\mu$m size (blow-out size), we
  arrive} at a rate of $\sim 10^{14}\g \s^{-1}$ \citep[also
see][]{tp07}. So for high $\alpha$ ($\alpha \geq 0.1$), the efficiency
of gas production is almost comparable to the dust removal rate. We
then have the intriguing possibility that dust in the $\beta$\,Pic
disk may not be {\rr all} blown away but is {\rr partially} accreted
back to the star {\rr in the form of gas. This possibility was also
  considered by \citet{zuckerman} for the bright debris disk of 49 Ceti.}

{\w In the super-solar case, we expect to see the gas
  being predominantly channelled into the central star.} Such an
accretion could potentially provide an explanation for the X-ray
emission observed from $\beta$\,Pic: the required accretion rate to
explain the X-ray emission
is 
$6.6\times10^{13}$ -- $6.6\times10^{15}$\,g\,s$^{-1}$ \citep{hem05},
compatible with the above production rates for high $\alpha$'s. {\w
  This gas, moreover, should be highly enriched in C and O and
  relatively depleted in other metals. This prediction can be tested
  using data like those from \citet{bou02}.}


\subsection{Carbon and Oxygen Ice-Lines}

Our result that C and O have to be produced at solar or even
super-solar abundance with respect to other elements has interesting
implications for the environments of proto-planetary disks. If the gas
originates from grains $\sim {\rr 100}$\,AU away from $\beta$\,Pic, then C
and O would have to be of at least solar abundance {\w in these
  grains}. This contrasts {\w with both the asteroid belt bodies and
  the more distant cometary bodies in the solar system.}

In the proto-solar nebula, the ice-line for water ice (of condensation
temperature 135\,K) is calculated to lie at $\sim 1$\,AU, while that
for CO ice (of condensation temperature 20\,K) is located at
10--100\,AU, depending on disk geometry \citep{chi97}.  The flatter
the disk geometry, the closer {\rr in} the ice line. 
Solid bodies formed between the two ice-lines would have been rich in
water but carbon poor, with densities close to 1\,g\,cm$^{-3}$.  This
is indeed observed for the satellites of Jupiter and Saturn \citep[see
review by][]{wongetal08}, and for smaller Kuiper belt objects
\citep{grundy,stansberry}.\footnote{Surprisingly, large Kuiper-belt
  objects like Pluto and Eris are too dense ($\sim$2\,g\,cm$^{-3}$) to
  be water enriched \citep{Null,sicardy}.}  Comets are also known to
be significantly depleted in carbon relative to water ice, independent
of whether they arrive from the Kuiper belt or the Oort cloud
\citep{MummaCharnley11}. This indicates that the carbon ice-line lies
outside $\sim 40$ AU in the solar system. 

Scaling from the solar model, with a stellar luminosity of $8.7
L_\odot$ for $\beta$\,Pic, the ice-line for CO should {\rr have lied}
around $30$--$300$\,AU. Here, our study indicate that the carbon
ice-line should {\rr have lied} inward of ${\rr \sim100}$\,AU, or,
the primordial disk is not very flared.

\subsection{Future Refinement}

In the present paper, we have described a local model which assumes
that gas is produced at 100\,AU, where most of the dust is located,
and compute how the radiation force and accretion differentially act
on the gas to evolve the abundances.
In reality, gas is observed to extend
{\rr from} at least 13\,AU from the central star \citep{bra04} {\rr to
  hundreds of AU}. If the gas is produced where most of the dust is
located, the question becomes how the gas is transported inwards. One
possibility is accretion triggered by MRI. To model this, however,
requires a stratified model. Another assumption of our analytical
model is that \ion{C}{2} braking is efficient already from start,
resulting in a constant drift velocity $u_{\rm x}$ and a constant
production rate $S_{\rm x}$ for all elements ``x''. A more realistic
scenario would be that the gas builds up over time with \ion{C}{2}
braking becoming efficient only after the C density has reached a
sufficient level.  A more refined model should thus take into account
both the spatial extension and the evolution of the gas,
{\rr to} provide more precise abundance predictions for various
elements,
{\rr yielding} stronger constraints on the origin of the gas around
$\beta$\,Pic.

\section{Summary}
\label{sec:summary}

In this paper, we investigate the unusual elemental abundance of the $\beta$\,Pic gas
disk. 
We find that to attain the observed C/O rich gas using our model, the
C and O gas must be produced at either solar or super-solar abundance
compared to other elements; the sub-solar case is excluded. In the
case of solar abundance, a low gas viscosity of $\alpha<10^{-3}$ is
required, as well as a gas production rate of
$\sim10^{11}$\,g\,s$^{-1}$.  Such a rate is achievable in the ECG
model \citep{cze07} where gas is produced by high velocity
micro-meteorites bombarding and sublimating the debris disk grains.
In this scenario, the currently observed overabundance of C and O
results from preferential depletion of other metallic elements by
radiative acceleration.

In the case of super-solar production of C and O, a high gas viscosity
is required ($\alpha \sim 0.1$), as well as a high production rate of
$\sim 10^{13}$\,g\,s$^{-1}$. Such a rate, as well as the exotic
production pattern, can be satisfied by photo-desorption if the recent
study on the photo-desorption of CO stands \citep{obe07}. The mass in
the $\beta$\,Pic dust disk may be partially ground down to micro
meteorites that are pushed out by radiation pressure, partly outgassed
and accreted into the star. The latter part may be significant.  As
this gas accretes into the central star, it may provide a natural
explanation for the observed x-ray emission from $\beta$\,Pic, unusual
for an A-star.

Detailed comparison between our model and the observed abundance
pattern leads us to prefer the super-solar case, but only by a small
margin. Our preference is informed by the fact that the column density
of Si is closer to (albeit still 10 times larger than) the observed
value in the super-solar case. Moreover, we argue that the gas disk
should support the magnetorotational instability which yields a
healthy viscosity. We propose the following tests to determine the
dominant {\rr gas production} mechanism:

\begin{itemize}

\item A re-analysis of the Al II absorption data of
  \citet{lag98}. Since Al should not be preferentially depleted
  by radiation pressure, its abundance ratio relative to C and O will
  inform us of the production pattern.

\item Analysis of UV and x-ray data like those in \citet{bou02}
  and \citet{hem05}. If the x-ray
  does arise from gas accretion, as opposed to chromospheric activity,
  we can use it to determine the gas accretion rate, as well as the
  chemical abundances.

\item The radial profile of the gas disk. If the outgassing occurs at
  ${\rr \sim100}$\,AU and most of the gas accretes into the star, we expect to see
  gas at smaller radii. This is a particularly exciting aspect because
  the giant planet, $\beta$\,Pic b \citep{lag10}, is believed
  to orbit at $\sim 10$ AU and may interact with the gas in
  interesting ways.

\end{itemize}

{\rr Our current model is limited in that the gas disk is approximated by a single zone at steady state. This, though unlikely to affect our basic conclusions, would need improvements to allow a more detailed comparison with observations.}


\acknowledgments {\rr We thank the anonymous referee, as well as
  Philippe Th{\'e}bault for the many constructive criticisms that
  improved the presentation of our paper.} This work was {\w initiated
  during} the International Summer Institute for Modeling in
Astrophysics (ISIMA) program held at the Kavli Institute for Astronomy
and Astrophysics, Peking University.  JWX was supported by the
National Natural Science Foundation of China (Nos.\,10833001 and
10925313), PhD training grant of China (20090091110002), and
Fundamental Research Funds for the Central Universities (1112020102),
and {he thanks Pascale Garaud for valuable discussions and
  suggestions}.  AB was supported by the Swedish National Space Board
(contract 84/08:1) and YW by the NSERC discovery grant.
 


\begin{thebibliography}{99}

\bibitem[{{Anders} \& {Grevesse}(1989)}]{and89}
{Anders}, E., \& {Grevesse}, N. 1989, \gca, 53, 197

\bibitem[{{Aumann}(1985)}]{aum85}
{Aumann}, H.~H. 1985, \pasp, 97, 885

\bibitem[{{Backman} \& {Paresce}(1993)}]{bac93}
{Backman}, D.~E., \& {Paresce}, F. 1993, in Protostars and Planets III, ed.
  E.~H. {Levy} \& J.~I. {Lunine}, 1253--1304

\bibitem[{{Bai} \& {Stone}(2011)}]{bai11}
{Bai}, X.-N., \& {Stone}, J.~M. 2011, \apj, 736, 144

\bibitem[{{Balbus}(2009)}]{bal09}
{Balbus}, S.~A. 2009, ArXiv e-prints

\bibitem[{{Balbus} \& {Hawley}(1991)}]{bal91}
{Balbus}, S.~A., \& {Hawley}, J.~F. 1991, \apj, 376, 214

\bibitem[{{Beust} {et~al.}(1989){Beust}, {Lagrange-Henri}, {Vidal-Madjar}, \&
  {Ferlet}}]{beu89}
{Beust}, H., {Lagrange-Henri}, A.~M., {Vidal-Madjar}, A., \& {Ferlet}, R. 1989,
  \aap, 223, 304

\bibitem[{{Bouret} {et~al.}(2002){Bouret}, {Deleuil}, {Lanz}, {Roberge},
  {Lecavelier des Etangs}, \& {Vidal-Madjar}}]{bou02}
{Bouret}, J.-C., {Deleuil}, M., {Lanz}, T., {Roberge}, A., {Lecavelier des
  Etangs}, A., \& {Vidal-Madjar}, A. 2002, \aap, 390, 1049

\bibitem[{{Brandeker}(2011)}]{bra11}
{Brandeker}, A. 2011, \apj, 729, 122

\bibitem[{{Brandeker} {et~al.}(2004){Brandeker}, {Liseau}, {Olofsson}, \&
  {Fridlund}}]{bra04}
{Brandeker}, A., {Liseau}, R., {Olofsson}, G., \& {Fridlund}, M. 2004, \aap,
  413, 681

\bibitem[{{Brandeker} {et~al.}(2012){Brandeker}, {Olofsson}, {Vandenbussche},
  {Acke}, {Barlow}, {Blommaert}, {Cohen}, {Dent}, C., \& J.}]{bra12}
{Brandeker}, A., {et~al.} 2012, subm. to \aap

\bibitem[{{Chen} {et~al.}(2007){Chen}, {Li}, {Bohac}, {Kim}, {Watson}, {van
  Cleve}, {Houck}, {Stapelfeldt}, {Werner}, {Rieke}, {Su}, {Marengo},
  {Backman}, {Beichman}, \& {Fazio}}]{che07}
{Chen}, C.~H., {et~al.} 2007, \apj, 666, 466

\bibitem[{{Chiang} \& {Goldreich}(1997)}]{chi97}
{Chiang}, E.~I., \& {Goldreich}, P. 1997, \apj, 490, 368

\bibitem[{{Crawford} {et~al.}(1994){Crawford}, {Spyromilio}, {Barlow}, {Diego},
  \& {Lagrange}}]{cra94}
{Crawford}, I.~A., {Spyromilio}, J., {Barlow}, M.~J., {Diego}, F., \&
  {Lagrange}, A.~M. 1994, \mnras, 266, L65

\bibitem[{{Czechowski} \& {Mann}(2007)}]{cze07}
{Czechowski}, A., \& {Mann}, I. 2007, \apj, 660, 1541

\bibitem[{{Fern{\'a}ndez} {et~al.}(2006){Fern{\'a}ndez}, {Brandeker}, \&
  {Wu}}]{fer06}
{Fern{\'a}ndez}, R., {Brandeker}, A., \& {Wu}, Y. 2006, \apj, 643, 509

\bibitem[{{Fleming} {et~al.}(2000){Fleming}, {Stone}, \& {Hawley}}]{fle95}
{Fleming}, T.~P., {Stone}, J.~M., \& {Hawley}, J.~F. 2000, \apj, 530, 464

\bibitem[{{Freudling} {et~al.}(1995){Freudling}, {Lagrange}, {Vidal-Madjar},
  {Ferlet}, \& {Forveille}}]{fre95}
{Freudling}, W., {Lagrange}, A.-M., {Vidal-Madjar}, A., {Ferlet}, R., \&
  {Forveille}, T. 1995, \aap, 301, 231

\bibitem[{{Grundy} {et~al.}(2007){Grundy}, {Stansberry}, {Noll}, {Stephens},
  {Trilling}, {Kern}, {Spencer}, {Cruikshank}, \& {Levison}}]{grundy}
{Grundy}, W.~M., {et~al.} 2007, \icarus, 191, 286

\bibitem[{{Hauschildt} {et~al.}(1999){Hauschildt}, {Allard}, \&
  {Baron}}]{hau99}
{Hauschildt}, P.~H., {Allard}, F., \& {Baron}, E. 1999, \apj, 512, 377

\bibitem[{{Hawley} {et~al.}(1995){Hawley}, {Gammie}, \& {Balbus}}]{haw95}
{Hawley}, J.~F., {Gammie}, C.~F., \& {Balbus}, S.~A. 1995, \apj, 440, 742

\bibitem[{{Hempel} {et~al.}(2005){Hempel}, {Robrade}, {Ness}, \&
  {Schmitt}}]{hem05}
{Hempel}, M., {Robrade}, J., {Ness}, J.-U., \& {Schmitt}, J.~H.~M.~M. 2005,
  \aap, 440, 727

\bibitem[{{Johnson}(2011)}]{joh11}
{Johnson}, R.~D., ed. 2011, {NIST Computational Chemistry Comparison and
  Benchmark Database NIST Standard Reference Database Number 101 Release 15bs}

\bibitem[{{Lagrange} {et~al.}(1995){Lagrange}, {Vidal-Madjar}, {Deleuil},
  {Emerich}, {Beust}, \& {Ferlet}}]{lag95}
{Lagrange}, A.~M., {Vidal-Madjar}, A., {Deleuil}, M., {Emerich}, C., {Beust},
  H., \& {Ferlet}, R. 1995, \aap, 296, 499

\bibitem[{{Lagrange} {et~al.}(1998){Lagrange}, {Beust}, {Mouillet}, {Deleuil},
  {Feldman}, {Ferlet}, {Hobbs}, {Lecavelier Des Etangs}, {Lissauer}, {McGrath},
  {McPhate}, {Spyromilio}, {Tobin}, \& {Vidal-Madjar}}]{lag98}
{Lagrange}, A.-M., {et~al.} 1998, \aap, 330, 1091

\bibitem[{{Lagrange} {et~al.}(2010){Lagrange}, {Bonnefoy}, {Chauvin}, {Apai},
  {Ehrenreich}, {Boccaletti}, {Gratadour}, {Rouan}, {Mouillet}, {Lacour}, \&
  {Kasper}}]{lag10}
---. 2010, Science, 329, 57

\bibitem[{{Lecavelier des Etangs} {et~al.}(2001){Lecavelier des Etangs},
  {Vidal-Madjar}, {Roberge}, {Feldman}, {Deleuil}, {Andr{\'e}}, {Blair},
  {Bouret}, {D{\'e}sert}, {Ferlet}, {Friedman}, {H{\'e}brard}, {Lemoine}, \&
  {Moos}}]{lec01}
{Lecavelier des Etangs}, A., {et~al.} 2001, \nat, 412, 706

\bibitem[{{Mentuch} {et~al.}(2008){Mentuch}, {Brandeker}, {van Kerkwijk},
  {Jayawardhana}, \& {Hauschildt}}]{men08}
{Mentuch}, E., {Brandeker}, A., {van Kerkwijk}, M.~H., {Jayawardhana}, R., \&
  {Hauschildt}, P.~H. 2008, \apj, 689, 1127

\bibitem[{{Mumma} \& {Charnley}(2011)}]{MummaCharnley11}
{Mumma}, M.~J., \& {Charnley}, S.~B. 2011, \araa, 49, 471

\bibitem[{{Null} {et~al.}(1993){Null}, {Owen}, \& {Synnott}}]{Null}
{Null}, G.~W., {Owen}, W.~M., \& {Synnott}, S.~P. 1993, \aj, 105, 2319

\bibitem[{{{\"O}berg} {et~al.}(2007){{\"O}berg}, {Fuchs}, {Awad}, {Fraser},
  {Schlemmer}, {van Dishoeck}, \& {Linnartz}}]{obe07}
{{\"O}berg}, K.~I., {Fuchs}, G.~W., {Awad}, Z., {Fraser}, H.~J., {Schlemmer},
  S., {van Dishoeck}, E.~F., \& {Linnartz}, H. 2007, \apjl, 662, L23

\bibitem[{{Olofsson} {et~al.}(2001){Olofsson}, {Liseau}, \&
  {Brandeker}}]{olo01}
{Olofsson}, G., {Liseau}, R., \& {Brandeker}, A. 2001, \apjl, 563, L77

\bibitem[{{Roberge} {et~al.}(2006){Roberge}, {Feldman}, {Weinberger},
  {Deleuil}, \& {Bouret}}]{rob06}
{Roberge}, A., {Feldman}, P.~D., {Weinberger}, A.~J., {Deleuil}, M., \&
  {Bouret}, J.-C. 2006, \nat, 441, 724

\bibitem[{{Shakura} \& {Sunyaev}(1973)}]{sha73}
{Shakura}, N.~I., \& {Sunyaev}, R.~A. 1973, \aap, 24, 337

\bibitem[{{Sicardy} {et~al.}(2011){Sicardy}, {Ortiz}, {Assafin}, {Jehin},
  {Maury}, {Lellouch}, {Hutton}, {Braga-Ribas}, {Colas}, {Hestroffer},
  {Lecacheux}, {Roques}, {Santos-Sanz}, {Widemann}, {Morales}, {Duffard},
  {Thirouin}, {Castro-Tirado}, {Jel{\'{\i}}nek}, {Kub{\'a}nek}, {Sota},
  {S{\'a}nchez-Ram{\'{\i}}rez}, {Andrei}, {Camargo}, {da Silva Neto}, {Gomes},
  {Martins}, {Gillon}, {Manfroid}, {Tozzi}, {Harlingten}, {Saravia}, {Behrend},
  {Mottola}, {Melendo}, {Peris}, {Fabregat}, {Madiedo}, {Cuesta}, {Eibe},
  {Ull{\'a}n}, {Organero}, {Pastor}, {de Los Reyes}, {Pedraz}, {Castro}, {de La
  Cueva}, {Muler}, {Steele}, {Cebri{\'a}n},
  {Monta{\~n}{\'e}s-Rodr{\'{\i}}guez}, {Oscoz}, {Weaver}, {Jacques}, {Corradi},
  {Santos}, {Reis}, {Milone}, {Emilio}, {Guti{\'e}rrez}, {V{\'a}zquez}, \&
  {Hern{\'a}ndez-Toledo}}]{sicardy}
{Sicardy}, B., {et~al.} 2011, \nat, 478, 493

\bibitem[{{Slettebak}(1975)}]{sle75}
{Slettebak}, A. 1975, \apj, 197, 137

\bibitem[{{Smith} \& {Terrile}(1984)}]{smi84}
{Smith}, B.~A., \& {Terrile}, R.~J. 1984, Science, 226, 1421

\bibitem[{{Stansberry} {et~al.}(2006){Stansberry}, {Grundy}, {Margot},
  {Cruikshank}, {Emery}, {Rieke}, \& {Trilling}}]{stansberry}
{Stansberry}, J.~A., {Grundy}, W.~M., {Margot}, J.~L., {Cruikshank}, D.~P.,
  {Emery}, J.~P., {Rieke}, G.~H., \& {Trilling}, D.~E. 2006, \apj, 643, 556

\bibitem[Th{\'e}bault 
\& Augereau(2007)]{tp07} Th{\'e}bault, P., \& Augereau, J.-C.\ 2007, \aap, 472, 169

\bibitem[{{van Leeuwen}(2007)}]{van07}
{van Leeuwen}, F. 2007, \aap, 474, 653

\bibitem[{{Vidal-Madjar} {et~al.}(1986){Vidal-Madjar}, {Ferlet}, {Hobbs},
  {Gry}, \& {Albert}}]{vid86}
{Vidal-Madjar}, A., {Ferlet}, R., {Hobbs}, L.~M., {Gry}, C., \& {Albert}, C.~E.
  1986, \aap, 167, 325

\bibitem[{{Vidal-Madjar} {et~al.}(1994){Vidal-Madjar}, {Lagrange-Henri},
  {Feldman}, {Beust}, {Lissauer}, {Deleuil}, {Ferlet}, {Gry}, {Hobbs},
  {McGrath}, {McPhate}, \& {Moos}}]{vid94}
{Vidal-Madjar}, A., {et~al.} 1994, \aap, 290, 245

\bibitem[{{Wong} {et~al.}(2008){Wong}, {Lunine}, {Atreya}, {Johnson},
  {Mahaffy}, {Owen}, \& {Encrenaz}}]{wongetal08}
{Wong}, M.~H., {Lunine}, J.~I., {Atreya}, S.~K., {Johnson}, T., {Mahaffy},
  P.~R., {Owen}, T.~C., \& {Encrenaz}, T. 2008, Reviews in Mineralogy and
  Geochemistry, 68, 219

\bibitem[{{Zagorovsky} {et~al.}(2010){Zagorovsky}, {Brandeker}, \&
  {Wu}}]{zag10}
{Zagorovsky}, K., {Brandeker}, A., \& {Wu}, Y. 2010, \apj, 720, 923

\bibitem[Zuckerman 
\& Song(2012)]{zuckerman} Zuckerman, B., \& Song, I.\ 2012, \apj, 758, 77 

\end{thebibliography}


\appendix
\section{A. \,   Equilibrium velocity due to neutral-ion collisions: $v_{\mathrm{neu},1}$}
Following \citet{beu89}, who studied the braking of ions by a neutral
gas, we here consider a neutral atom braked by an ion gas (consisting
of \ion{C}{2} at a number density of $N_{\mathrm{C\,II}}$).  The
average net momentum loss of the neutral atom in one collision is
$-m_{\mathrm{C}}{\bf v}$, where $m_{\mathrm{C}}$ is the mass of one C
atom and ${\bf v}$ is the neutral {\rr drift velocity relative to \ion{C}{2},
with $v$ its magnitude.}  On average, the collisions are then
equivalent to an effective force on the atom that can be expressed as
\begin{equation}
{\bf F}_{\mathrm{neu}}=- N_{\mathrm{C\,II}}\pi b_{\mathrm{ni}}^{2} m_{\mathrm{C}}v{\bf v}=-k \frac{v}{v_{\mathrm{cl}}} \,{\bf v}.
\end{equation}
Here 
\begin{equation}
b_{\mathrm{ni}}=\left(\frac{1}{4\pi\epsilon_{0}}\frac{4e^{2}P_{\mathrm{x}}} {\mu \, v_{\mathrm{cl}}^{2}}\right)^{1/4}
\end{equation}
is the largest impact parameter that can lead to a physical collision between the neutral and the ion, where $\epsilon_{0}$ is the permittivity of free space, $e$ is the charge of an electron, $P_{\mathrm{x}}$ is the polarizability of the neutral ``x'',
$\mu=m_{\mathrm{x}}m_{\mathrm{C}}/(m_{\mathrm{x}}+m_{\mathrm{C}})$ is the reduced mass,
$v_{\mathrm{cl}}$ is the neutral-ion collision velocity, and where
\begin{equation}
k=\pi m_{\mathrm{C}}\sqrt{\frac{4e^{2}P_{\mathrm{x}}}{4\pi\epsilon_{0}}\, \frac{N_{\mathrm{C\,II}}^{2}}{\mu}}.
\end{equation}
Two parts are contributing to the collisional velocity,
\begin{equation}
v_{\mathrm{cl}}\sim v_{\mathrm{s}}+v, 
\label{vcl}
\end{equation}
where $v_{\mathrm{s}}$ is the sound speed of \ion{C}{2} gas. The atom,   
on the other hand, is subject to the radiation force from the central star, which is 
\begin{equation}
{\bf F}_{\mathrm{rad}}\sim\beta \frac{GM_{\star}m_{\mathrm{x}}} {r^{2}}.
\end{equation} 
Equating ${\bf F}_{\mathrm{neu}}$ to ${\bf F}_{\mathrm{rad}}$, the equilibrium velocity can be solved as 
\begin{equation}
v_{\rm neu,1}=\frac{F_{\mathrm{rad}}+\sqrt{F_{\mathrm{rad}}^{2}+4k\,v_{\mathrm{s}}F_{\mathrm{rad}}} } {2k}.
\label{v1a}
\end{equation}
This is the limiting drift velocity of a radiatively driven atom braked by neutral-ion collisions.
If $v\gg v_{\mathrm{s}}$ (an assumption that here holds for most elements), then equation \ref{vcl} is reduced to $v_{\mathrm{cl}}\sim v$, leading to 
\begin{eqnarray}
v_{\rm neu,{1}} & \sim & F_{\mathrm{rad}}/k  \ \ \ \ \ {\rm if} \ \ \ \ v\gg v_{\mathrm{s}}, \nonumber \\
& =  & \frac{\beta}{\pi}\left(\frac{m_{\mathrm{x}}}{m_{\mathrm{C}}}\right) \left(\frac{GM_{\star}}{r^{2}}\right) \left(\frac{4e^{2}P_{\mathrm{x}}}{4\pi\epsilon_{0}} \, \frac{N_{\mathrm{C\,II}}^{2}}{\mu}\right)^{-1/2},\nonumber \\
& \sim & 80 \beta \left( \frac{m_{\mathrm{x}}}{m_{\mathrm{C}}} \right) \left(\frac{m_{\mathrm{x}}m_{\mathrm{C}}} {m_{\mathrm{p}}(m_{\mathrm{x}}+m_{\mathrm{C}})}\right)^{1/2}  \left( \frac{N_{\mathrm{C\,II}}}{100\,\rm cm^{-3}}\right)^{-1} \left(\frac{r}{100\,\rm AU} \right)^{-2}   \left( \frac{M_{\star}}{\rm M_{\odot}}\right) \left( \frac{P_{\mathrm{x}}}{10\, \rm \AA^{3}}\right)^{-1/2}\,\rm cm\,s^{-1},
\label{vneu1-app}
\end{eqnarray}
which is the case considered by \citet{beu89}. \\ \\

\section{B. \,   Equilibrium velocity due to ion-ion collisions: $v_{\mathrm{ion}}$}
Here we study the case of a tracer ion moving in \ion{C}{2} gas.
Following \citet{beu89}, according to the classical theory of Coulomb scattering (here referred to as ``collision''), the average net momentum loss of a tracer ion colliding with a C ion is 
\begin{eqnarray}
\delta p= -m_{\mathrm{C}}{\bf v}(\cos\chi -1),
\label{dp}
\end{eqnarray}
where $\chi$ is the deflection angle given by 
\begin{eqnarray}
\tan \frac{\chi}{2}= \frac{1}{4\pi\epsilon_{0}}\frac{e^{2}}{\mu} \frac{1}{bv_{\mathrm{cl}}^{2}},
\label{chi}
\end{eqnarray}
where $b$ is the impact parameter. $v_{\mathrm{cl}}$ is the ion-ion impact velocity as given by equation \ref{vcl}. In \citet{beu89}, they implicitly ignore $v_{s}$ and take $v_{\mathrm{cl}}\sim v$. In the case we consider here, however, the tracer ion would be efficiently braked by the \ion{C}{2}, implying $v\ll  v_{\mathrm{s}}$ and $v_{\mathrm{cl}}\sim v_{\mathrm{s}}$. 

The equivalent force due to ion-ion collisions can be expressed as 
\begin{eqnarray}
{\bf F}_{ion} &  =&v N_{\mathrm{C\,II}} \int_{0}^{b_{\mathrm{max}}} (\delta p)2\pi b \, \, {\rm d}b \nonumber \\
& = & 2 N_{\mathrm{C\,II}} \pi b_{\mathrm{ii}}^{2} \, \ln\left(\frac{\mu^{2}}{m_{\mathrm{C}}^{2}}\frac{\lambda_{\mathrm{D}}^{2}}{b_{\mathrm{ii}}^{2}} +1 \right) \, \frac{m_{\mathrm{C}}^{2}}{\mu^{2}}m_{\mathrm{C}}v \, {\bf v},   \ \ \ \ \ \ \ {\rm  if} \ \ \ v_{\mathrm{cl}}= v_{\mathrm{s}} \nonumber \\
& \sim & 4 N_{\mathrm{C\,II}} \pi b_{\mathrm{ii}}^{2} \, \ln\left(\frac{\lambda_{\mathrm{D}}}{b_{\mathrm{ii}}} \right) \, \left(1+\frac{m_{\mathrm{C}}}{m_{\mathrm{x}}}\right) m_{\mathrm{C}}v \, {\bf v}.
\label{fion}
\end{eqnarray} 
Here $b_{\mathrm{max}}$ is the maximum acceptable value of the impact parameter, which can be taken as the Debye length
\begin{eqnarray}
\lambda_{\mathrm{D}}\sim \sqrt{\frac{\epsilon_{0}k_{\mathrm{B}}T_{e}} {e^{2}n_{e}} }\sim\sqrt{\frac{\epsilon_{0}k_{\mathrm{B}}T_{\mathrm{disk}}} {e^{2}N_{\mathrm{C\,II}}} },  \ \ \ \ {\rm if} \ \ \ n_{e}\sim N_{\mathrm{C\,II}} \ \ \ {\rm and} \ \ \  T_{e}\sim T_{\mathrm{disk}},
\label{lambdaD}
\end{eqnarray}
where $k_{\mathrm{B}}$ is the Boltzmann constant, $T_{e}$ and $T_{\mathrm{disk}}$ are the temperatures of the electrons and the disk, respectively, and $n_{e}$ is the number density of electrons. If we assume C to be the major donator of electrons, then we expect $n_{e}\sim N_{\mathrm{C\,II}}$.
$b_{\mathrm{ii}}$ is a characteristic impact parameter for the field ion-ion collision (obtained by solving equation \ref{chi} for $\chi=90^{\circ}$ and substituting $\mu$ with $m_{\mathrm{C}}$):
\begin{eqnarray}
b_{\mathrm{ii}}= \frac{1}{4\pi\epsilon_{0}}\frac{e^{2}}{m_{\mathrm{C}}} \frac{1}{v_{s}^{2}} \sim \frac{1}{4\pi\epsilon_{0}} \frac{e^{2}}{k_{\mathrm{B}}T_{\mathrm{disk}}},
\label{bii}
\end{eqnarray}
thus 
\begin{eqnarray}
\frac{\lambda_{\mathrm{D}}}{b_{\mathrm{ii}}}= 4\pi \, \epsilon^{3/2} \, e^{-3} \,  k_{\mathrm{B}}^{3/2} \, T_{\mathrm{disk}}^{3/2} \, N_{\mathrm{C\,II}}^{-1/2}.
\label{bii}
\end{eqnarray}

Equating ${\bf F}_{ion}$ with ${\bf F}_{rad}$, we find the equilibrium velocity to be
\begin{eqnarray}
v_{\rm ion} & =  &\left[\left(\frac{\beta}{4\, {\rm ln} (\lambda_{\mathrm{D}}/b_{\mathrm{ii}})}\right)\, \left(\frac{1}{\pi b_{\mathrm{ii}}^{2} r N_{\mathrm{C\,II}}} \right ) \left(\frac{m_{\mathrm{x}}^{2}} {m_{\mathrm{C}}(m_{\mathrm{C}}+m_{\mathrm{x}})}\right)  \right]^{1/2}   \left (\frac{GM_{\star}}{r}\right)^{1/2} \nonumber \\
& \sim & 3.7 \beta \left[\frac{m_{\mathrm{x}}^{2}} {m_{\mathrm{C}}(m_{\mathrm{C}}+m_{\mathrm{x}})}\right]^{1/2} \left(\frac{N_{\mathrm{C\,II}}}{100\, \rm cm^{-3}}\right)^{-1/2}  \left(\frac{r}{100\, \rm AU}\right)^{-1}   \left( \frac{M_{\star}}{\rm M_{\odot}}\right)^{1/2} \left(\frac{T_{\mathrm{disk}}}{100 \, \rm K}\right) \,\, {\rm cm\,s^{-1}}.
\end{eqnarray} \\

\end{document}